\documentclass[twocolumn]{aastex631}
\usepackage[utf8]{inputenc}
\usepackage{float}
\usepackage{amsmath, bm, bbm}
\usepackage{amssymb}
\usepackage{physics}
\usepackage{tikz}
\usepackage{enumitem}

\DeclareRobustCommand{\bbone}{\text{\usefont{U}{bbold}{m}{n}1}}
\usepackage{hyperref}
\usepackage{tikz}

\usepackage{color}

\begin{document}

\title{Beyond Gaussian Noise: A Generalized Approach to Likelihood Analysis with non-Gaussian Noise}

\author{Ronan Legin}\thanks{These authors contributed equally.}
\affiliation{Department of Physics, Université de Montréal, Montréal, Canada}
\affiliation{Ciela - Montreal Institute for Astrophysical Data Analysis and Machine Learning, Montréal, Canada}
\affiliation{Mila - Quebec Artificial Intelligence Institute, Montréal, Canada}

\author{Alexandre Adam}\thanks{These authors contributed equally.}
\affiliation{Department of Physics, Université de Montréal, Montréal, Canada}
\affiliation{Ciela - Montreal Institute for Astrophysical Data Analysis and Machine Learning, Montréal, Canada}
\affiliation{Mila - Quebec Artificial Intelligence Institute, Montréal, Canada}

\author{Yashar Hezaveh}
\affiliation{Department of Physics, Université de Montréal, Montréal, Canada}
\affiliation{Ciela - Montreal Institute for Astrophysical Data Analysis and Machine Learning, Montréal, Canada}
\affiliation{Mila - Quebec Artificial Intelligence Institute, Montréal, Canada}
\affiliation{Center for Computational Astrophysics, Flatiron Institute, 162 5th Avenue, 10010, New York, NY, USA}

\author{Laurence Perreault Levasseur}
\affiliation{Department of Physics, Université de Montréal, Montréal, Canada}
\affiliation{Ciela - Montreal Institute for Astrophysical Data Analysis and Machine Learning, Montréal, Canada}
\affiliation{Mila - Quebec Artificial Intelligence Institute, Montréal, Canada}
\affiliation{Center for Computational Astrophysics, Flatiron Institute, 162 5th Avenue, 10010, New York, NY, USA}

\begin{abstract}
Likelihood analysis is typically limited to normally distributed noise due to the difficulty of determining the probability density function of complex, high-dimensional, non-Gaussian, and anisotropic noise. This is a major limitation for precision measurements in many domains of science, including astrophysics, for example, for the  analysis of the Cosmic Microwave Background, gravitational waves, gravitational lensing, and exoplanets.
This work presents Score-based LIkelihood Characterization (SLIC), a framework that resolves this issue by building a data-driven noise model using a set of noise realizations from observations. We show that the approach produces unbiased and precise likelihoods even in the presence of highly non-Gaussian correlated and spatially varying noise. 
We use diffusion generative models to estimate the gradient of the probability density of noise with respect to data elements. In combination with the Jacobian of the physical model of the signal, we use Langevin sampling to produce independent samples from the unbiased likelihood. We demonstrate the effectiveness of the method using real data from the \textit{Hubble Space Telescope} and \textit{James Webb Space Telescope}. 
\end{abstract}

\keywords{
        Astronomy data modeling(1859) ---
        Astronomy data analysis(1858) ---
        Measurement error model(1946) ---
        Astronomy data modeling(1859) ---
        Algorithms(1883)
}

\section{Introduction}

The presence of measurement errors, or noise, in data leads to uncertainty in the inference of variables of interest. This uncertainty is quantified by the likelihood, which depends on the statistics of the measurement errors \citep{MacKay2003}. In laboratory and observational settings, additive noise is the most common form of measurement error. In this case, the observed signal, such as the values of pixels in a CCD image, is the sum of the true underlying signal and a stochastic noise term. Evaluating the likelihood requires knowledge of the noise model or the probability density function of noise realizations.

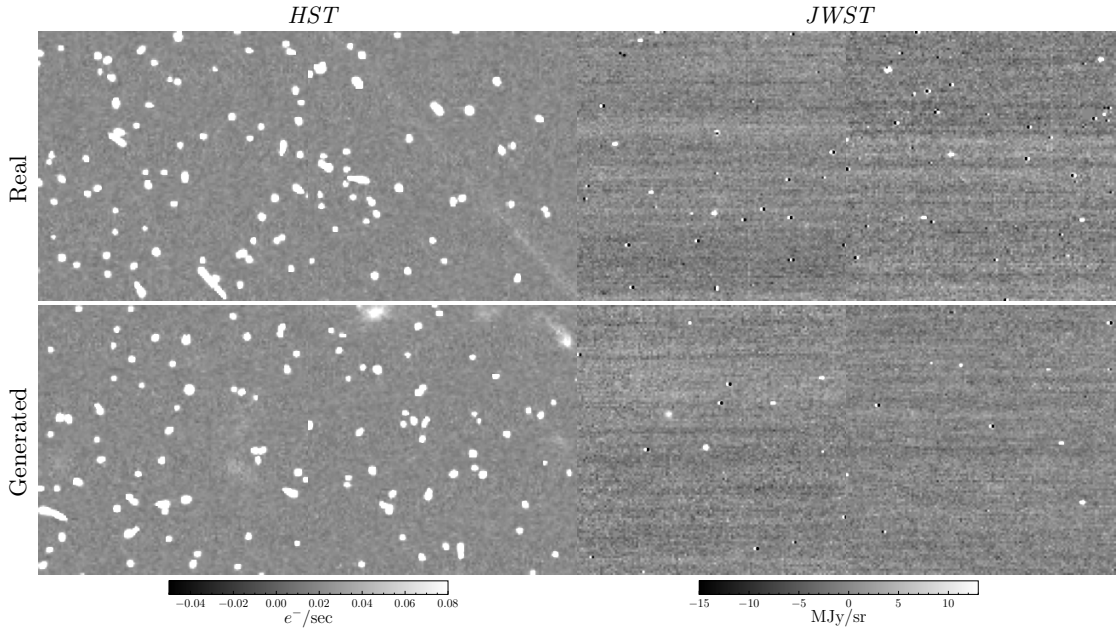
\begin{figure*}[ht!]
\centering
\begin{tikzpicture}
    \node at (0, 0) {\includegraphics[width=0.8\textwidth]{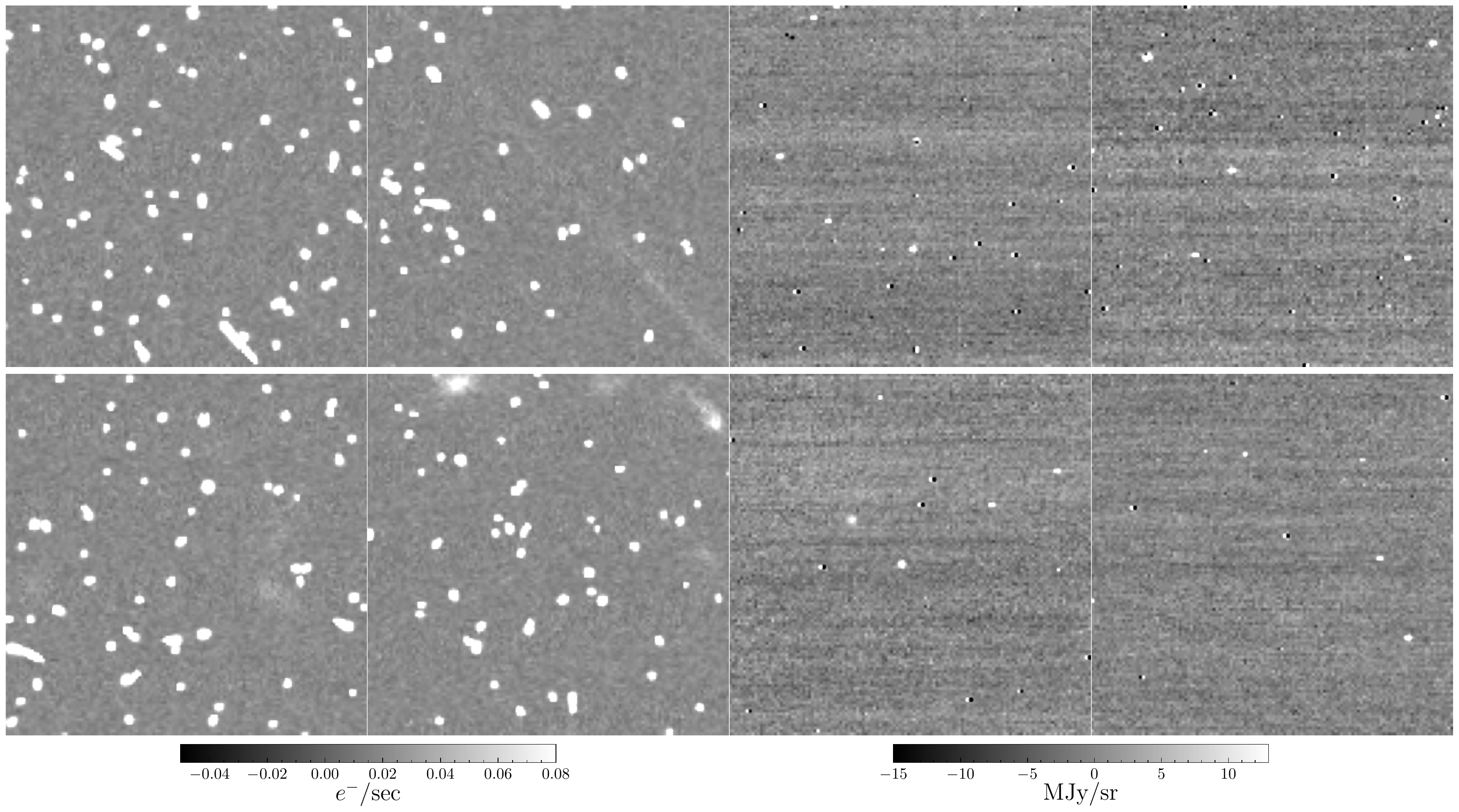}};
    \node at (-3.5, 4.15) {\strut \textit{HST}};
    \node at (3.5, 4.15) {\strut \textit{JWST}};
    \node[rotate=90, align=center] at (-7.4, 2) {\strut Real};
    \node[rotate=90, align=center] at (-7.4, -1.5) {\strut Generated};
\end{tikzpicture}
\caption{\textit{Top row}: Examples of noise patches from the datasets. \textit{Bottom row}: Examples of noise patches generated using the score models used in this work. These examples are generated by solving the reverse-time 
variance-exploding stochastic differential equation with 1000 steps of the discretized Euler-Maruyama solver.}
\label{fig:samples}
\end{figure*}

Errors in data can be caused by various physical processes, resulting in different forms of noise such as readout noise and thermal (dark) noise \citep[e.g., ][]{Somiya2006, Gilliland2011, Konnik2014, Harper2018, Chen2020, Abbott2020, Schlawin_2020}. In some cases, noise is the result of the additive effect of many microscopic stochastic processes. In such scenarios, the central limit theorem states that the probability distribution of noise can be well approximated by a normal distribution \citep[e.g., ][]{Kwak2017}. For example, thermal noise in CCD pixels is caused by many small-scale quantum processes, resulting in noise that is almost perfectly normally distributed, allowing for simple likelihood evaluations. This well-motivated approximation and the resulting simplicity have led to the widespread adoption of normal noise models in likelihoods across many fields of science \citep[e.g.,][]{Carron2013, Fichet2016, Wang2019}.

In practice, however, the statistics of noise often deviate significantly from a simple normal distribution and could be highly correlated, anistropic, and non-Gaussian. Examples include time-correlated read-out noise in detectors \citep[e.g., ][]{Payeur:22}, cosmic rays impacting CCD pixels \citep[e.g., ][]{Giardino2019, Miles2021}, anistropic (e.g., non-uniform scanning of the sky) and correlated (e.g., $1/f$) noise in CMB maps \citep[e.g., ][]{Madhavacheril}, and abrupt noise \textit{glitches} in gravitational wave signals due to disturbances from the environment or instrument \citep[e.g., ][]{Bose2016, Powell2016, Zackay2021, Yu2022}. 

Traditionally, the approach to deal with these scenarios has been to adopt an approximate normal noise model for the likelihood (and accept the potential biases that result), model and remove specific forms of noise \citep[e.g., cosmic rays removal from CCD images, ][]{Windhorst1994, Pych2004, Zhang2020}, masking a subset of data with significant non-Gaussian noise, or avoiding using likelihood analysis altogether. 

These sub-optimal approaches are driven by the fact that the probability density of complex, high-dimensional, non-Gaussian noise is not known and is intractable with closed-form expressions. 
An alternative approach is to use simulation-based inference \citep{Cranmer:20} if a generative model of noise is available or to build a correct noise model from simulations or observations of noise realizations \citep[e.g., ][]{Hahn:19}. 

In this work, we aim to do the latter and build a noise model using a set of observed noise realizations in high dimensions (e.g., pixel space) and show that this can produce unbiased likelihoods even for noise with highly non-Gaussian statistics. We would like to clarify that the focus of this paper is on non-Gaussianity in likelihoods arising from non-Gaussian noise, not from non-linearity in physical models. The latter could simply be dealt with using MCMC samplers.


We focus on the ubiquitous case of additive noise and assume that the noise realizations used for learning the noise model are either simulated or real data in the absence of the signal of interest (e.g., regions in a CCD image that do not contain the signal of interest).

Our framework, Score-based LIkelihood Characterization (SLIC), uses score-based diffusion generative models \citep{Song2019, Ho2020, Song2020} to estimate the gradient of the probability density of noise with respect to image pixels and show that, combined with the Jacobian of the physical model of the signal, one can use Langevin sampling to produce independent samples of the likelihood.

In section \ref{sec:framework} we describe SLIC and explain how to produce samples from a likelihood in presence of non-Gaussian noise, in section \ref{sec:results}, we describe our experiments with non-Gaussian noise and present the results of the proposed framework and compare it against Gaussian pseudo-likelihoods. In section \ref{sec:discussionconclusion} we discuss our results and conclude.

\section{Framework}\label{sec:framework}

Standard likelihood sampling methods (e.g., Metropolis-Hasting Markov chain Monte Carlo, MCMC) require the probability density of the likelihood to be evaluated for any given input \citep{Speagle2019}. Langevin sampling, however, offers an alternative possibility to only use the gradient of the logarithm of the probability density (the score function) to produce samples from a probability distribution\footnote{Note that Langevin sampling is also an MCMC procedure.}.
Here, we first summarize the standard approach of training a score-based model on realizations of noise and then explain how to use the learned score in combination with the Jacobian of the physical model to sample the likelihood.

\subsection{Learning the noise model}
\label{sec:scoremodel}
We assume that we have a dataset of $n$ images of noise, where each image has $m$ elements (i.e. the number of pixels), $D_{\mathrm{Training}} = \{\mathbf{x}_1, \mathbf{x}_2, ..., \mathbf{x}_n\}$. We construct this training set with i.i.d. realization of noise from an underlying true noise probability density distribution, $Q_{\mathrm{True}}(\mathbf{x})$. 
Here, $\mathbf{x}$ refers to any $m$-dimensional point (i.e. an image) in the data space. When evaluated on a particular data, e.g., $\mathbf{x}_0$, this function informs us of how likely it is that $\mathbf{x}_0$ is a realization of noise. 
Our goal is to build a model for the probability density function of noise, $Q(\mathbf{x})$.

In the context of score-based models, instead of learning $Q$ directly, we learn the gradient of the logarithm of $Q$ with respect to its variable $\mathbf{x}$, written as $\mathbf{s}(\mathbf{x}_0) = \partial \log Q(\mathbf{x}_0) / \partial \mathbf{x}$. We refer the reader to \cite{Song2020} for a comprehensive description of how to learn $\mathbf{s}(\mathbf{x})$. In summary, the method involves training a machine learning model using the score matching technique \citep{hyvarinen2005, Vincent2011, Alain2014} to estimate the score. This is achieved by perturbing the data with the forward process of a variance-exploding stochastic differential equation. The goal of this procedure is to help the network learn the score over the entire domain of the distribution by perturbing the data with Gaussian noise (distinct from the noise distribution we are aiming to learn) --- in essence, smoothing the target distribution with a temperature.

Typically, the model is a neural network. For example, in the case of image data, a neural network that has the same dimensionality for the input and output images \citep[e.g., a UNet, ][]{Ronneberger2015} can be used. The network takes an input image, $\mathbf{x}$, and produces an output image, $\mathbf{s}(\mathbf{x})$, where each pixel in the output image represents an approximation to the gradient of the logarithm of the probability density with respect to the corresponding pixel in the input image.

\subsection{Sampling the Parameters of the Physical Model}

Our goal is to generate samples from the parameters of the physical model, $\eta \in \mathbb{R}^{l}$, with the probability given by the likelihood. The data-generating process can be written as:
\begin{equation}\label{eq:data_generating_process} 
\begin{aligned}
        \mathbf{x} &= \mathbf{M}(\mathbf{\eta}) + \mathbf{N} 
\end{aligned}
\end{equation} 
where $\mathbf{M}(\mathbf{\eta})$ encodes the signal (the physical model) as a function of the parameters of interest $\mathbf{\eta}$ and $\mathbf{N}$ is a vector of additive noise. A particular observation could be written as $\mathbf{x_{\mathrm{O}}} = \mathbf{M}(\mathbf{\eta}_{\mathrm{True}}) + \mathbf{N}_0$, where $\mathbf{N}_0$ is the true (but unknown) realization of noise.

Since here we are only considering non-stochastic physical models, the likelihood $P(\mathbf{x_{\mathrm{O}}} \vert \eta)$ can be written as
\begin{equation}\label{eq:MainEquation} 
\begin{aligned}
        P(\mathbf{x_{\mathrm{O}}} \vert \eta) = Q(\mathbf{x_{\mathrm{O}}}-\mathbf{M}(\eta) )
\end{aligned}
\end{equation} 
where $Q$ is the probability density of noise. 

If we can compute the score of the likelihood, $\grad_{\eta} \log Q$, we can then sample from the underlying likelihood distribution, $Q(\mathbf{x_{\mathrm{O}}}-\mathbf{M}(\eta))$, using an overdamped Langevin diffusion process:
\begin{equation}\label{eq:langevin} 
\begin{aligned}
        d \mathbf{\eta} &= \grad_{\eta} \log Q(\mathbf{x_{\mathrm{O}}}-\mathbf{M}(\eta) ) dt +  \sqrt{2} d\mathbf{w}
\end{aligned}
\end{equation} 
where $\mathbf{w}$ is an $l$-dimensional Brownian process (for $l$-dimensional variable $\eta$). This stochastic differential equation (SDE) can be discretized by the well-known Euler-Maruyama scheme to yield the unadjusted Langevin sampling algorithm
\begin{equation}\label{eq:langevin} 
\begin{aligned}
        \mathbf{\eta}_{t+1} &= \mathbf{\eta}_t + \tau \grad_{\eta} \log Q(\mathbf{x_{\mathrm{O}}}-\mathbf{M}(\eta) ) +  \sqrt{2 \tau} \mathbf{\xi}_t
\end{aligned}
\end{equation} 
where $\tau$ is a step size and $\mathbf{\xi}_t \sim \mathcal{N}(0, \bbone)$ is a random variable drawn from an $l$-dimensional multivariate normal distribution.

Starting from an arbitrary $\mathbf{\eta}_0$, the walker diffuses with the help of the random diffusion term (the third term on the r.h.s. of equation \ref{eq:langevin}) while being guided by the gradient of the distribution of interest (the second term on the r.h.s.) towards high probability regions.

After a sufficiently long random walk (the burn-in phase) and under sufficient regularity conditions, the walker produces a random sample from the distribution of interest. This is consistent with the fact that the distribution of walkers driven by this SDE will obey the Fokker-Planck equation \citep[e.g., ][]{Lai:22}. To obtain multiple independent samples, one could either continue the chain for a significantly longer period of time (past its auto-correlation time) or simply start the sampling procedure from the beginning and perform a new walk. The convergence properties of this sampling method are studied in detail in \citet{Roberts1996}.

\begin{figure*}[ht!]
        \centering
		\includegraphics[width=0.9\linewidth]{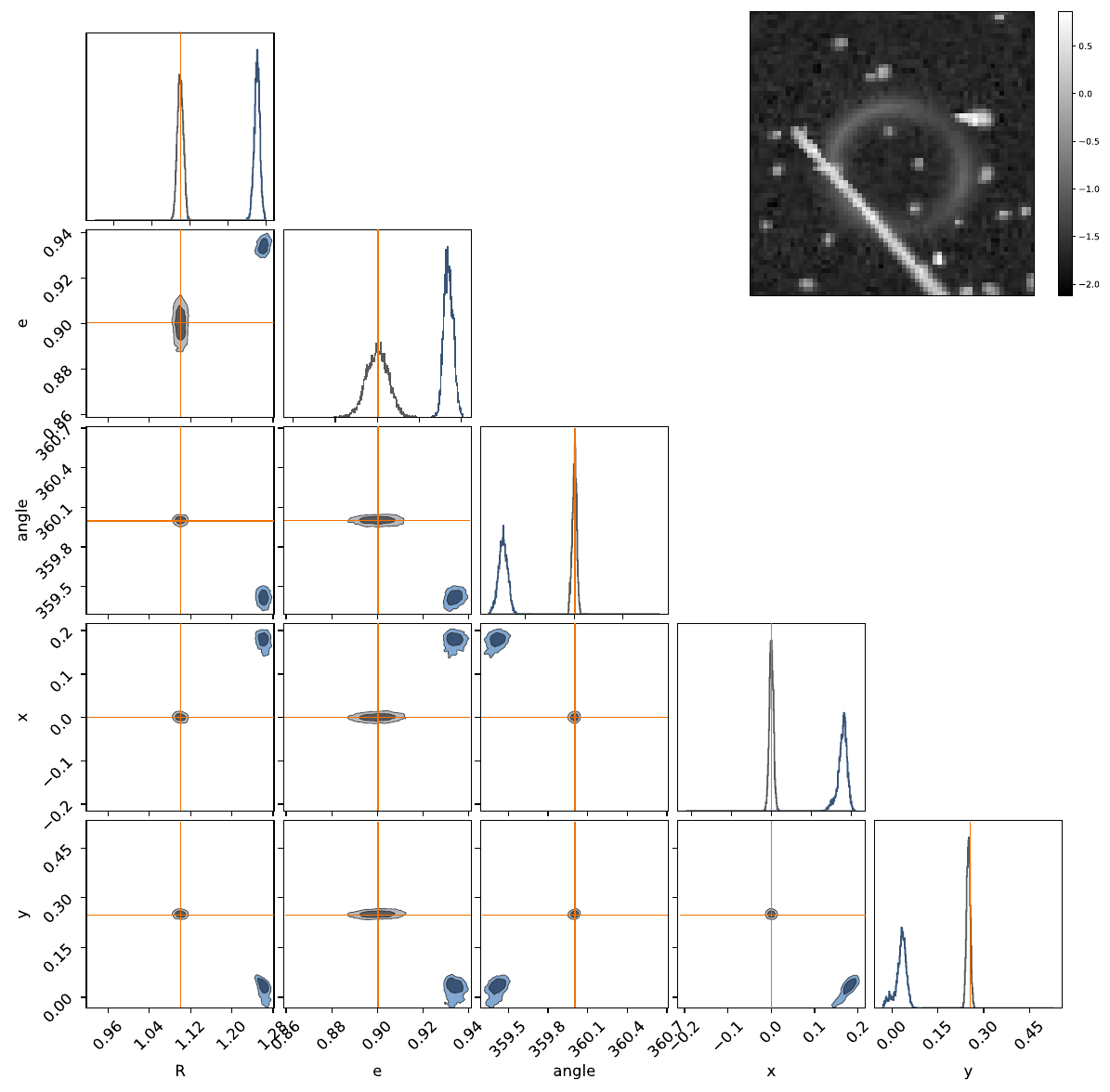}  
        \caption{The analysis of a gravitational lensing signal with noise taken from \textit{HST} data. The data is shown on the top panel (in $\log_{10}$). The gray contours show samples generated with SLIC and the dark blue contours show biased results when a Gaussian likelihood is sampled with an MCMC initialized at the true parameter values.}
        \label{fig:hst_corner}
\end{figure*}

\begin{figure*}[ht!]
        \centering
		\includegraphics[width=0.9\linewidth]{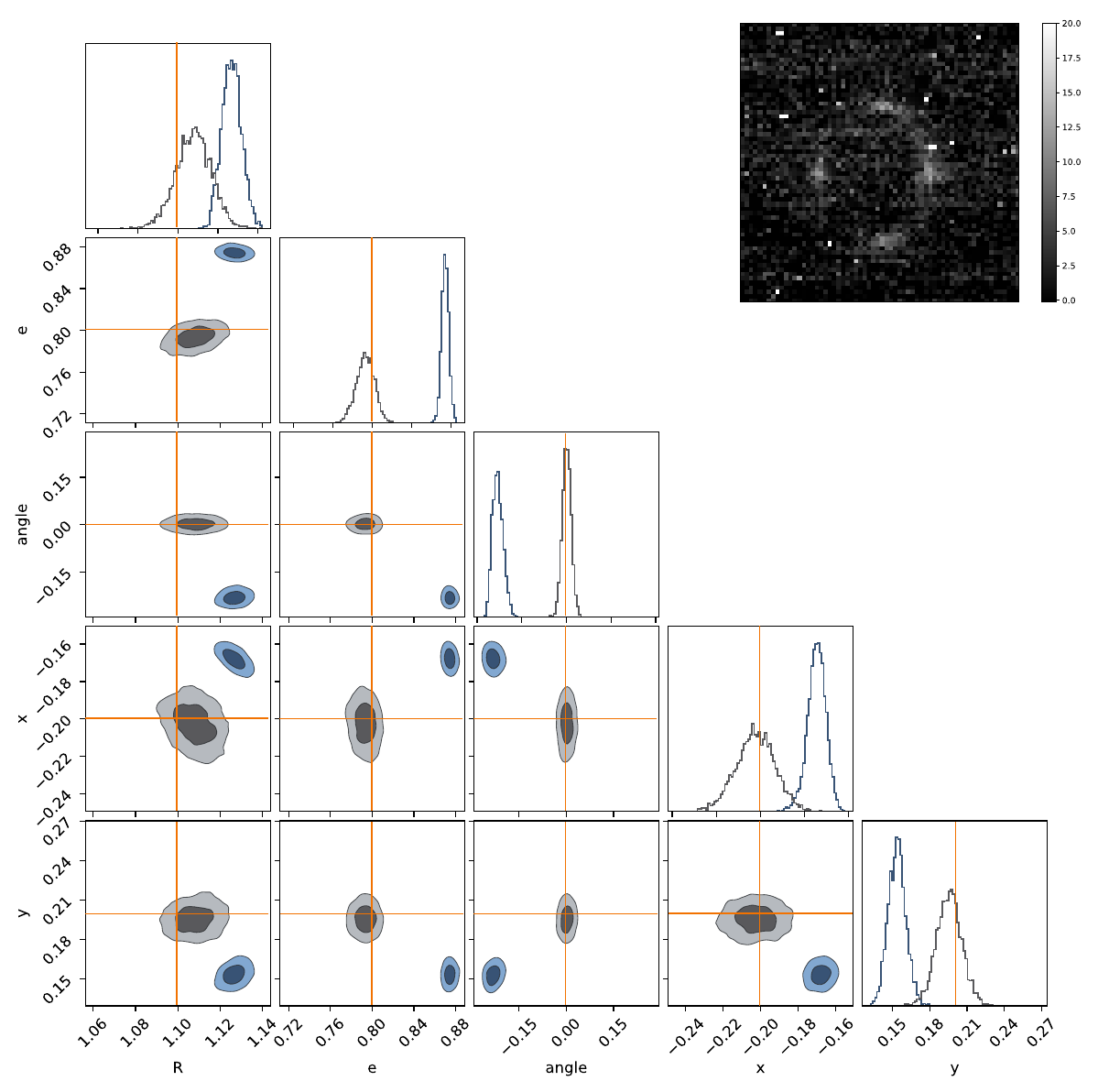}  
        \caption{Same as Figure \ref{fig:hst_corner} but performed on \textit{JWST} data (image is in linear intensity units).}
        \label{fig:jwst_corner}
\end{figure*}

To get the score of the likelihood in the parameters of the model, we use the chain rule to write:
\begin{equation}\label{eq:MainEquation} 
\begin{aligned}
\grad_{\eta} \log Q &=
\grad_{\mathbf{x}} \log Q
 \cdot  \grad_{\mathbf{\eta}} \mathbf{x}
\end{aligned}
\end{equation} 
Here, $\grad_{\mathbf{x}} \log Q$ is the score of the noise model with respect to the pixel values, which is the variable learned by the neural network score function and $\grad_{\mathbf{\eta}} \mathbf{x}$ is the gradient of the pixel values with respect to the variables of interest $\eta$. Since the pixel values only change when the model $\mathbf{M}(\mathbf{\eta})$ changes, this is equivalent to $\partial \mathbf{M} / \partial \eta$, which is the Jacobian of the physical model. 

To summarize, the procedure to generate samples from the likelihood with SLIC involves:
\begin{enumerate}
\item Learning a score model for the distribution of noise in the data space, $\mathbf{s}(\mathbf{x}) = \grad_{\mathbf{x}} \log Q(\mathbf{x})$.
\item  Calculating the Jacobian of the physical model, $\partial \mathbf{M} / \partial \eta$. 
\item Using the Euler-Marayuma SDE solver to produce samples from the likelihood:
\begin{equation}\label{eq:MainEquation} 
\hspace{-15pt}
\begin{aligned}
        \mathbf{\eta}_{i+1} &= \mathbf{\eta}_{i} + 
        \tau \,  \grad_{\mathbf{x}} \log Q(\mathbf{x_{\mathrm{o}}}-\mathbf{M}(\eta)) \grad_{\eta} M(\mathbf{\eta}_i) + \sqrt{2 \tau} \xi
\end{aligned}
\end{equation} 
\end{enumerate}

\section{Experiments and Results}
\label{sec:results}

In this section, we present the results of our experiments to model and sample likelihoods in the presence of non-Gaussian noise using SLIC. We use \textit{Hubble Space Telescope} (\textit{HST}) and \textit{James Webb Space Telescope} (\textit{JWST}) imaging data, publicly available at \url{https://archive.stsci.edu/} and \url{https://hla.stsci.edu/}, to test the proposed framework. We also compare the samples from our likelihood to those generated from a Gaussian likelihood with an MCMC sampler.

For \textit{JWST}, we use five $2048 \times 2048$ stage-2 calibrated exposure-based images to produce empty noise cutouts. The images were taken by the \textit{JWST} Near Infrared Camera (NIRCam) under program ID 11631 using the F444W filter with an exposure time of 96.631 seconds. In total, we produce a set of approximately 800 $128 \times 128$ \textit{JWST} images to train the score model. For \textit{HST}, we use around 50 different snapshots mainly consisting of background noise and cosmic rays, and produce a second training set of approximately 150000 $64 \times 64$ \textit{HST} cutouts.

Once the score model is trained using the procedure described in section \ref{sec:scoremodel}, we can visually assess the quality of its performance by generating new realizations of noise using the method described in \citet{Song2020}. Figure \ref{fig:samples} shows examples of real \textit{HST} and \textit{JWST} noise along with noise realizations generated from the score model.

To produce simulated observations with real non-Gaussian \textit{HST} and \textit{JWST} noise, we inject the signal of a strongly lensed galaxy into our real noise samples. Note that the noise sample was not seen during the training of the score model. The signal is produced by distorting the image of a S\'{e}rsic background source by the gravity of a Singular Isothermal Ellipsoid (SIE) foreground structure. Our gravitational lensing model is implemented in \texttt{JAX} and is fully differentiable, allowing us to easily calculate the Jacobian of the model. Our goal is to infer the parameters of the foreground model (mass, ellipticity, orientation angle, position $x$ and $y$). 

We then use Langevin dynamics to produce 2000 samples from the likelihood using equation \ref{eq:langevin}, with a step size of $\tau=10^{-5}$. 

To compare our results with a Gaussian likelihood, we approximate the variance of the noise over regions without extremely bright pixels and produce samples from the Gaussian likelihood using an MCMC sampler \citep[\texttt{emcee}, ][]{Foreman2013}. The samplers are initiated from the true parameters.

Figures \ref{fig:hst_corner} and \ref{fig:jwst_corner} show the results for data generated with \textit{HST} and \textit{JWST} respectively. The figures compare samples produced by SLIC (gray contours) with those from a likelihood for Gaussian noise (blue contours), showing that SLIC produces accurate likelihoods for highly non-Gaussian noise.

\begin{figure}[ht!]
        \centering
		\includegraphics[width=\linewidth]{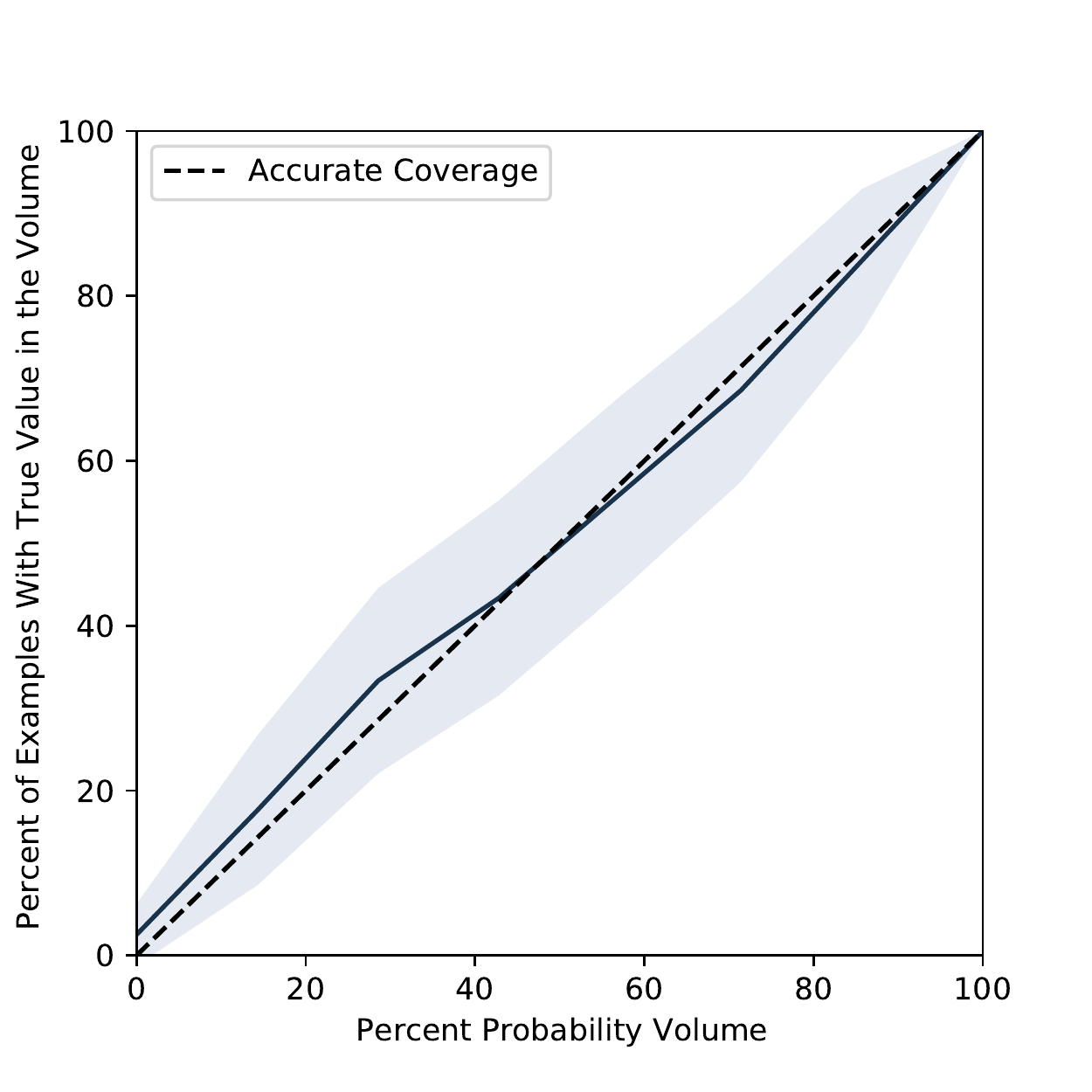}  
        \caption{The coverage probability calculated by estimating the SLIC likelihoods for 200 different realizations of \textit{HST} noise for a signal similar to that of figure \ref{fig:hst_corner}. The $3\sigma$ error region (shaded) is calculated using jacknifing. The trend of following the dashed diagonal line shows that the likelihood is accurate (unbiased).}
        \label{fig:coverage}
\end{figure}

Figure \ref{fig:coverage} shows the coverage probability calculated using 200 different realizations of \textit{HST} noise for a signal similar to that of figure \ref{fig:hst_corner}. In every experiment, we use the same ground truth parameters but add different realizations of noise to the signal and obtain the corresponding SLIC likelihood. The coverage probabilities are calculated using the method proposed by \citet{Lemos:23}.

\section{Discussion and conclusion}
\label{sec:discussionconclusion}
Our experiments show that by modeling the empirical distribution of noise and having access to differentiable physical models, it is possible to build accurate likelihood models for highly non-Gaussian noise, resulting in accurate inference without the need for major approximations. In multiple other experiments with real and simulated noise with extreme non-Gaussianity, we observe that SLIC consistently produces unbiased likelihoods, while likelihoods that assume Gaussian noise are highly biased (the MCMC chains for the Gaussian likelihood are started at the ground truth parameters and are fully converged). 

As shown in figures \ref{fig:hst_corner} and \ref{fig:jwst_corner}, the procedure provides likelihoods with high accuracy while maintaining the precision and the information content of the data. Note that the framework is not only limited to instrumental noise and can also be used to treat any other sources of additive errors, including additive contamination by astrophysical background and foreground sources. Additionally, while we demonstrated the method using gravitational lensing as an example, the framework is general and can be used for the inference of any differentiable physical model.

The proposed framework also offers the advantage of utilizing the learned noise models for generating new noise realizations, as shown in Figure \ref{fig:samples}. These simulations can be used to train other inference models (e.g., in the context of simulation-based inference) or for conducting forecasting experiments.

SLIC provides a powerful approach to model the entire distribution of noise without parameterizing the noise distribution and being limited to a particular sub-space. For example, even in the presence of Gaussian noise, using a normal likelihood with a fixed covariance matrix can result in a bias, since the true covariance is not known and an estimate of it is used. For accurate inference, one should marginalize over the unknown parameter of this model, i.e. the covariance matrix \citep{Sellentin:16}. In SLIC, however, if different noise realizations with different covariance matrices are included in the training data, the model will effectively consider them in the likelihood to the degree with which the residuals are likely to be generated from each covariance, resulting in marginalization over different noise scenarios. Therefore, if there are multiple noise characteristics with an instrument (e.g., different exposure lengths), they should all be included as a single training dataset. The distribution of the noise statistics in the training data is then the prior over the noise statistics. This suggests that for precision data analysis, SLIC could provide an advantage even in the presence of Gaussian noise, since the likelihood of Gaussian noise is, in effect, not Gaussian \citep{Sellentin:16,Hahn:19}.

While in this work we used an unadjusted Langevin method to produce samples from the score of the likelihood, it is also possible to use Hamiltonian MCMC to sample the non-Gaussian likelihoods \citep[e.g., see][]{Remy:22}.

A subtle point to consider for the sampling of the likelihood is that the score model is trained with progressively increasing levels of diffusive noise, parameterized by a temperature. The true distribution of the data (which is of interest for our purpose) corresponds to the lowest temperature distribution. However, this distribution does not have support in regions that are significantly far from the noise manifold. This happens when the starting point in the sampling chains is significantly far from the true parameter, producing large residuals which are the input to the score model. In this case, to guide the model to the true noise manifold, the score model should be used at a higher temperature and the temperature should be decreased progressively.

While in our experiments the number of parameters for inference is relatively small ($l=5$), for the inference of high dimensional variables, the Jacobian of the model is a very large matrix of size $m \times l$, where $m$ is the dimension of the data $\in \mathbb{R}^m$, and $l$ is the number of parameters of interest $\eta \in \mathbb{R}^l$, requiring a significant amount of memory to store. To circumvent this, a possibility would be to use the Vector-Jacobian product (VJP). 
This would allow the treatment the Jacobian matrix as a function that takes as input a cotangent vector --- a vector that belongs to the image of $M(\eta)$ like $\grad_\mathbf{x} \log Q (\mathbf{x})$ --- 
and returns the gradient $\grad_\eta \log Q$ as its output. 
Note that, in this case, the total computational and memory cost of the VJP would be of the same order as evaluating the forward model $M(\eta)$ and the score model $\grad_{\mathbf{x}} \log Q$. 

We conclude by summarizing that given a set of possibly highly non-Gaussian noise realizations, SLIC provides a framework to use a score-based generative model to characterize the likelihood and to produce samples from it. This allows accurate analysis of highly non-Gaussian likelihoods without approximations that result in biases or loss of precision. The method could contribute to significantly improved likelihood analysis in precision fields including the CMB, gravitational waves, gravitational lensing, and exoplanets.

\section*{Acknowledgements}
This research was made possible by a generous donation by Eric and Wendy Schmidt with the recommendation of the Schmidt Futures Foundation.

We thank Fran\c{c}ois Lanusse and Gil Holder for reading the manuscript and providing constructive comments.

The work is in part supported by computational resources provided by Calcul Quebec and the Digital Research Alliance of Canada. Y.H. and L.P. acknowledge support from the National Sciences and Engineering Council of Canada grant RGPIN-2020-05102, the Fonds de recherche du Québec grant 2022-NC-301305 and 300397, and the Canada Research Chairs Program. R.L. acknowledges support from the Centre for Research in Astrophysics of Quebec and the hospitality of the Flatiron Institute. 

\bibliography{bibliography}

\end{document}